# Building and Governing AI Systems: Advancing Social Workers' Roles across the Technology Industry, Human Service Organizations, and Policy Institutions

Nari Yoo, Daphne Watkins, Brian Perron, Jessica Kane, and Matthew Smith

School of Social Work, University of Michigan, Ann Arbor, MI, USA

Corresponding author: Nari Yoo, School of Social Work, University of Michigan, Ann Arbor, MI 48109, USA. Email: nariyoo@umich.edu. ORCID: 0000-0002-8467-7048

## Abstract

Artificial intelligence is moving the technology sector into domains social work has long served, including crisis response, mental health care, benefits administration, vocational rehabilitation, and child welfare. Social workers meet technology teams as users of their tools, as subjects in their datasets, and as first responders to what those systems deploy, yet they study these systems from outside the settings where the decisions are made. This paper introduces the standard roles on a technology product team and the decisions each one controls, reviews the disciplines around AI-era technology together with the social work scholarship that meets each, and identifies five groups of technology decision roles social workers can hold across the technology industry, human service organizations, and policy institutions, spanning product, governance, organizational technology leadership, grantee collaboration, and policy work. Product management is our primary example. We set the nine competencies of the 2022 Educational Policy and Accreditation Standards against the tasks that role performs to show where MSW training already teaches the method the job asks for. We then specify the technical understanding these roles require, extend the profession's technology ethics from use to deployment, and close with a research agenda for the profession's technology workforce.

**Keywords**: artificial intelligence; social work profession; product management; social work education; EPAS competencies; human service organizations; AI governance; AI literacy

## Introduction

Artificial intelligence (AI) has moved from pilot projects to operating infrastructure in the public systems where social workers practice (Code for America, 2026; Office of the Federal Chief Information Officer, 2025; U.S. Government Accountability Office, 2025). Federal agencies reported 3,611 cases of AI use in 2025, with the United States Department of Health and Human Services (DHHS) reporting more use cases than any other agency (Office of the Federal Chief Information Officer, 2025). An independent audit of 11 agencies reported cases of AI use nearly doubling between 2023 and 2024 (U.S. Government Accountability Office, 2025). At the state level, a cross-state assessment found AI already embedded in benefits document processing, fraud detection, and eligibility workflows (Code for America, 2026).

Much of the investment in AI lands in domains where social workers already deliver services. That investment is financial and contractual, moving through procurement contracts between government agencies and AI vendors (Anthropic, 2025; Code for America, 2026), federal grant dollars passed from federal agencies to States (Alms, 2026), and deployments that payers and managed care organizations fund inside their own operations (Axelrod et al., 2025). Maryland contracted with the AI developer Anthropic to deploy an assistant that will help residents apply for SNAP, Medicaid, cash assistance, and WIC, and will help caseworkers verify documents and validate eligibility in a program that manually processes more than 150,000 documents each month (Anthropic, 2025). Medicaid managed care organizations already apply AI to risk stratification and utilization management, with minimal public documentation of methods or regulatory oversight (Axelrod et al., 2025). Meanwhile, DHHS's Administration for Children and Families committed $6 million to States interested in expanding the piloting of AI-

based predictive risk analytics[1] in child welfare (Alms, 2026). In addition, crisis response now runs through a national call, text, and chat platform, the 988 Suicide and Crisis Lifeline, which routed approximately 19.1 million contacts between July 2022 and September 2025 (Saunders, 2025; U.S. Government Accountability Office, 2026a). Each of these deployments is an AI system in the sense the NIST AI Risk Management Framework uses, an engineered or machine-based system that can, for a given set of objectives, generate outputs such as predictions, recommendations, or decisions influencing real or virtual environments (Tabassi, 2023). In benefits administration, child welfare, and crisis response, the objectives are set over case, client, or applicant data, and a caseworker, an applicant, or an automated process acts on the output.

Alongside government procurement, private philanthropy is financing AI adoption in the nonprofit and community-based organizations that deliver safety-net and human services. As technology firms' influence has grown and expectations of corporate social responsibility have risen, their philanthropy has grown into a distinct, market-oriented form, a marketization of giving built on the technology sector's own values of innovation, entrepreneurialism, and a focus on financial and performance metrics (Manning et al., 2020). For instance, the Ballmer Group, the Gates Foundation, and their partner funders committed $8.5 million in 2025 to AI-enabled tools for access to public benefits, including food assistance, Medicaid, housing, and unemployment services (Center for Civic Futures, 2025). The OpenAI Foundation (2026) has directed unrestricted grants to more than 200 community-based organizations. A workforce has grown up around these systems as well; in a 2024 survey of more than 670 professionals across 45 countries, 77% of organizations reported working on AI governance (IAPP & Credo AI, 2025). The same expansion has already produced failures that reach clients, including the

[1] Predictive risk models are statistical models that score a case on the estimated probability of a future adverse event and rank cases for worker attention on that score (Gillingham, 2016; Saxena et al., 2020).

Department of Veterans Affairs' retirement of an AI tool used in suicide prevention without documentation of lessons learned (U.S. Government Accountability Office, 2026b).

Social workers are implicated in these systems as the workers who operate them and as the professionals whose clients are assessed by them, and they are rarely present in the decisions that set how a system behaves. Those decisions determine which problem the system addresses, what data it is built and run on, what output it returns, what threshold triggers an action, and who reviews it before release. The absence is not explained by a small workforce. Licensed clinical social workers, at 221,791 in 2020, constituted the largest single category of licensed behavioral health specialists, ahead of licensed professional counselors, psychologists, and psychiatrists (U.S. Government Accountability Office, 2022). The published literature on how digital mental health tools are built, however, rarely registers that presence, and where clinical expertise appears in it, that expertise is consulted and does not decide. In a systematic review of 26 papers reporting on 24 digital mental health interventions, clinicians were the second most commonly involved stakeholder group after end users, appearing in 17 of 24 studies, yet end users in none of the interventions reviewed reached the highest, decision-maker level of involvement (Brotherdale et al., 2024). A mapping review of 30 e-mental health studies found designers explicitly named in only 8 of them and software development companies in 14, with the companies appearing only at the development step (Vial et al., 2022). Neither review counts social workers as a group of their own, since the systematic review records clinicians and the mapping review records health professionals and experts (Brotherdale et al., 2024; Vial et al., 2022). One included study names them, placing clinical psychologists, social workers, psychiatrists, and occupational therapists in pre- and post-design focus groups (Hetrick et al., 2018). The same pattern appears beyond mental health studies. For example, a human-centered review of the algorithms used in the U.S. child welfare system found their development dominated by risk-modeling objectives, with caseworkers' knowledge and case narratives largely left out (Saxena et al., 2020). An analysis of predictive risk modeling in child protection described these systems as black boxes that cannot be interrogated by the practitioners responsible for acting on them (Gillingham, 2016).

Social work scholarship has responded to the technology sector's growth with a literature of its own, beginning more than a decade ago with the profession's Grand Challenge to Harness Technology for Social Good (Berzin & Singer, 2015). Recent systematic and integrative reviews have 1) cataloged AI applications across the profession's practice domains (Garkisch & Goldkind, 2024; Li et al., 2026; Massey et al., 2026); 2) documented practitioners' and students' training needs and apprehensions (Kapur et al., 2026; Pandya, 2026); 3) designed tutorials to introduce the underlying AI techniques to research audiences (B. E. Perron, Luan, et al., 2025; B. E. Perron, Rivenburgh, et al., 2025), and 4) offered early guidance on utilizing generative AI in research and education (Patton et al., 2023; Reamer, 2023; Rodriguez et al., 2024; Victor et al., 2023). A specialty series on AI and social work (Victor & Herrenkohl, 2025) has added two further pieces. Ahn et al. (2025) specified the AI literacy that the social work profession's core competencies now require, and Lee et al. (2026) proposed an ethical framework for social work researchers' own use of AI. The claim that the profession belongs inside design work has been made before. Writing in an Association for Computing Machinery magazine, Patton (2020) argued that social work thinking should inform user experience and AI design, and that the profession’s methods for working with vulnerable populations are the ones such design requires. Less is known about the occupational positions surrounding AI-related roles themselves, meaning the roles inside technology companies, human service organizations, and policy institutions where decisions about AI and technology systems are made.

We set out those positions for a social work research audience, arguing that as the technology sector expands into the domains social work serves, social workers should hold decision roles across it, in product, governance, organizational technology leadership, grantee collaboration, and policy positions. The paper describes the technology product team role by role, locates the disciplines a social worker entering these settings will meet, develops the case for the product manager role against the nine competencies of the 2022 Educational Policy and Accreditation Standards, and specifies the understanding and the ethics these decisions require.

## The Technology Team and Its Decisions

Technology products are built and maintained by standing teams that work in short, repeating development cycles, sequencing their work through an ordered plan of what will be built over coming quarters, and often through objectives and key results that define what will count as success (Cagan, 2019; Maglyas et al., 2013). Two features of this arrangement bear directly on human service settings that adopt such products. First, because development is cyclical, software is released in versions and revised continuously, so the tool a caseworker uses on any given day reflects decisions that are still in progress. Second, a commercial vendor's build plan aggregates demand across its entire customer base, so a single agency's needs enter that plan as one input among many; a requested change that conflicts with the vendor's priorities may never be scheduled. For a profession whose clients live with the results of these decisions, the practical question is which member of the team holds which decision, and who is responsible for the overall outcome of those decisions?

In the software engineering literature, the *product manager* is defined less by technical tasks than by responsibility for the product's direction. Interview-based research with practicing software product managers identified four variants of the role, which differ along three dimensions, namely authority over the product, influence on collaboration, and access to resources (Maglyas et al., 2013). In every variant, the product manager is responsible for deciding which user problem the team addresses next, in what order features are built under fixed time and staffing, and whether the version that was released produced the intended result. Frameworks written for practicing product managers describe the role the same way across settings, with that responsibility carried through influence over a cross-functional team and not through formal authority over the engineers and designers on it, and with its competencies organized into user and domain knowledge, discovery and delivery techniques, and stakeholder management (Cagan, 2019). How much authority the position actually carries varies with how formally the organization has defined it (Ebert, 2007).

*Software engineers* hold the decisions about feasibility and the cost of change. Software engineering economics established that correcting a requirement grows substantially more expensive the later in development it is caught (Boehm, 1981), which gives engineering judgment continuous influence over what product managers can ask for and when. Case study evidence from large-scale development shows requirements formed by product roles being narrowed, deferred, or lost in the communication gaps between those roles and the engineers implementing them (Bjarnason et al., 2011). Surveys of software practitioners find architectural choices to be the leading source of technical debt, the accumulated cost of expedient engineering decisions that later constrains what a product can become (Ernst et al., 2015). A feature that engineering judges costly to build frequently leaves that plan without any other role having formally decided against it.

Design scholarship characterizes the *product designer's* contribution as problem framing. Buchanan (1992) described design as the discipline that takes up ill-defined, "wicked" problems and gives them workable form, a description that fits the role's daily authority over the sequence of screens a user encounters, the options presented at each step, and the defaults, meaning the option a screen has already selected before anyone touches it. Defaults decide outcomes at scale because most people accept them, as the gap between consent rates in opt-in and opt-out organ donation systems shows (Johnson & Goldstein, 2004), and a hurried caseworker or a client in crisis is among the least likely to change one. In products used for benefits applications or crisis support, such as the assistant Maryland residents use to apply for SNAP, Medicaid, and cash assistance (Anthropic, 2025) and the text and chat channels of the 988 Lifeline (Saunders, 2025), decisions about screen order, available options, and defaults determine which questions are asked first and how much a user must already know for the system to work for them. That same authority over defaults and option ordering can be turned against users; analyses of dark patterns, meaning interface choices that steer people toward outcomes favoring the vendor (Gray et al., 2018), document the practice at commercial scale across thousands of sites (Mathur et al., 2019).

The *user experience researcher* supplies the team's systematic evidence about users. Standard treatments of the role describe systematic investigation of users' needs, behavior, and contexts through interviews, field observation, and usability testing, translated into findings the team treats as grounds for decisions (Goodman et al., 2012), though those findings often fail to reach the decisions they were meant to inform (Wixon, 2003). The role's decisions concern whose experience is studied, which questions are asked, and which findings reach the plan; when the users consulted do not include the clients a human service agency serves, the product's evidence base inherits that omission, a bias design-justice scholarship attributes to methods that settle in advance whose needs count as the user's (Costanza-Chock, 2020).

*Data scientists* occupy a set of positions that empirical study has differentiated into several distinct types, from insight providers who inform team decisions to modeling specialists who build the predictive components of the product itself (M. Kim et al., 2016). The role decides which metric defines success, how online controlled experiments comparing product versions are designed and interpreted (Kohavi et al., 2009), and what target a predictive model optimizes. When a benefits platform ranks applications by predicted risk, or a crisis service routes contacts by predicted severity, the optimization target and the threshold at which the system acts were set by this role, and both fall directly on the populations social work serves.

*Trust and safety* names the function responsible for harm operations, the detection of and response to abuse, exploitation, self-harm content, and fraud on a platform. Scholars and practitioners of the field describe it as young and historically inward-facing, with a professional association, a dedicated peer-reviewed journal, and a shared terminology developing only in recent years (Badiei et al., 2023). Scholarship on content moderation treats these decisions as constitutive of what a platform is (Gillespie, 2018), and ethnography of the moderation workforce documents the escalation thresholds and human toll built into the work (Roberts, 2019). Its decisions, including what content is removed, which accounts are restricted, and when a case escalates to a human responder, constitute the platform-side counterpart of crisis protocols in social work settings.

*Platform policy and legal* teams write the rules that trust and safety operations enforce. Klonick (2018), drawing on access to the teams that built the content-governance systems at Facebook, Twitter, and YouTube, described a rule-drafting and case-adjudication process she compared to a common-law system, operating largely outside public view while governing expression for billions of users; other scholars describe these terms-of-service regimes as a private legal system that governs users with little accountability (Suzor, 2019); a systems-thinking account describes the governance itself as upstream rule-setting and automated enforcement operating at scale (Douek, 2022). For products that hold client information, this layer also sets the terms of data collection, retention, and sharing, decisions the social work profession's confidentiality standards would treat as consequential in any practice setting.

A newer set of governance roles operates alongside these product roles in many larger organizations. Responsible AI functions review products against fairness and accountability commitments, though interviews with the staff who do this work find their influence contingent on organizational support (Rakova et al., 2021), and red teams test models adversarially before release, meaning they attack a model deliberately to find what it will do wrong, a practice now documented at methodological length (Ganguli et al., 2022). The most institutionalized clinical version of this work lies outside the technology industry. NHS England's clinical risk management standards for health information technology require both manufacturers and deploying organizations to designate a Clinical Safety Officer, a senior clinician holding current professional registration who is accountable for the system's safety documentation (NHS England, 2025a, 2025b). These governance structures should not be mistaken for stable career territory, however, as several companies dissolved or sharply reduced their AI ethics and trust and safety teams during the 2022-2023 layoff wave (Field & Vanian, 2023).

**The Disciplines around AI-Era Technology**

The fields surrounding AI-era technology differ in their degree of institutionalization. Some are disciplines in the full institutional sense, with degree programs, departments, dedicated journals, and faculty lines; others are research communities organized around a shared concern that has not yet acquired that apparatus. The distinction parallels the account of roles, in that fields that have hardened into disciplines offer established training routes into the settings named here, whereas communities still unsettled about their own names offer routes that are more open and less defined. Figure 1 shows each field's connection to social work; Table 1 adds each field's principal research community.

**Figure 1**. Disciplines and fields at the intersection of social work and technology

Note. Solid circles are discipline-grade fields with departments and degree programs; dashed circles are research communities not yet institutionalized as disciplines. The line from the center names a contact point where social work scholarship engages that field.

Human-computer interaction (HCI), a computing subfield with degree programs and department-level institutional presence, asks how people use computing systems and how those systems should be designed around human capacities and contexts. Several of its methods parallel methods the social work profession already trains. Contextual inquiry builds understanding through observation and interviewing conducted while users do real work (Holtzblatt & Beyer, 1997), which is the procedure of a person-in-environment assessment, where the worker observes and interviews the client in the setting the problem occurs in (Council on Social Work Education, 2022). The participatory design tradition, which developed out of Scandinavian workplace democracy projects, holds that the people affected by a system should share authority in designing it (Bødker & Kyng, 2018), which is the commitment community-based participatory research makes when it gives community partners authority over the research question, the data, and the interpretation (Israel et al., 1998). Social work scholars have begun publishing within and alongside the field. For instance, a participatory design study of generative AI tools for social service practitioners was co-authored by social work faculty (Tan et al., 2025), and social work researchers developed an AI-assisted co-design process for statewide housing procedures (Ferguson et al., 2026).

Artificial intelligence and machine learning research, itself a computer science subfield, supplies the technical core the surrounding fields apply and critique, developing the models that classify inputs, predict outcomes, and generate text, images, and other content. Social work's methods literature has begun translating this core for its own research audience, with tutorial articles introducing word embeddings and the programming interfaces through which large language models are used (B. E. Perron, Luan, et al., 2025; B. E. Perron, Rivenburgh, et al., 2025). Data science, now organized into degree programs and schools of its own, concerns the principles and processes for extracting useful patterns from data and connecting those patterns to decisions (Provost & Fawcett, 2013). Text-mining models to identify domestic violence in child welfare case narratives reached better than 90% accuracy against human coders, with inter-rater reliability above a Fleiss kappa of .80 (Victor et al., 2021), and machine learning models predicting which youth are most likely to exit care without a permanent placement have been developed by social work scholars from administrative child welfare data (Ahn et al., 2021).

In 2009, computational social science was announced as a field arguing that digital behavioral data permit the study of social phenomena at scale (Lazer et al., 2009). Eleven years later, many of the same authors judged the field's methodological growth to have outpaced its infrastructure for data access, ethics review, and training (Lazer et al., 2020). Its methods operate on the unstructured and large-scale data that sociology, political science, and their neighbors increasingly use, from social media text and images to administrative records. Social work's contact with the field runs through the same material. Patton and colleagues paired qualitative coding led by social work researchers with natural language processing to interpret social media posts from gang-involved youth (Patton et al., 2016), and later work employed formerly gang-involved young people as domain experts within the annotation process itself (Frey et al., 2020). Social work researchers have also applied these methods to administrative and workforce records (Thomas et al., 2024; Yoo et al., 2025). The term “computational social work” has since entered the literature as a methodological program (Rodriguez & Lee, 2026).

Information science, institutionalized in the iSchools, asks how information is organized and used, and the social work profession's longest-standing technology tradition speaks this field's language. An international association for human service information technology applications, husITa, held its first international conference in Birmingham, England, in 1987 (Ballantyne, 2017). Its affiliated journal, now the *Journal of Technology in Human Services*, launched in 1985 under a social work founding editor (Ballantyne, 2017), and Parker-Oliver and Demiris (2006) argued in *Social Work* for establishing social work informatics as a named specialty. Science and technology studies, for its part, examines how technologies are produced, adopted, and embedded in social and political arrangements. Critical scholarship widely read in social work belongs to this broader tradition, including Eubanks's (2018) case studies of automated eligibility, predictive risk modeling, and coordinated entry in U.S. public services. The social work profession's own critical voices have extended it to AI deployment in statutory practice (Garrett, 2025) and to the political economy that an integrative review found the social work profession's AI literature had largely overlooked (Massey et al., 2026).

Several adjacent fields border the same work. Health informatics, organized through its own professional association and degree programs, includes nursing informatics. Digital health is a large and active field of its own, spanning digital mental health interventions, mobile health applications, and wearable sensing. It borders social work's clinical and mental health practice most directly, and much of the social work profession's intervention-facing technology work appears in its co-design and effectiveness research (Brotherdale et al., 2024; Vial et al., 2022). Implementation science examines how evidence-supported practices move into routine service settings, a question continuous with how human service organizations adopt technology. Robotics and human-robot interaction studies embody systems, with care robotics as a point of contact with the populations social work serves, and the learning sciences and educational technology adjoin the social work profession's growing literature on AI literacy and curriculum (Ahn et al., 2025; Rodriguez et al., 2024).

Two further concentrations of AI work are better described as emerging research communities. Responsible AI scholarship, whose principal peer-reviewed venue has been the Association for Computing Machinery (ACM) conference on Fairness, Accountability, and Transparency (FAccT) since 2018, studies how automated systems distribute benefit and harm and how they can be held accountable. A graduate-level training on fairness evaluation methods now exists (Barocas et al., 2023), and research with industry practitioners documents responsible AI as a staffed organizational function inside technology companies (Rakova et al., 2021). Social work has entered this community from two directions, with social work faculty publishing a fairness audit of a machine learning model used to allocate child welfare preventive services (Ahn et al., 2024) and the social work profession's ethics literature specifying the questions such audits should answer (Reamer, 2023). Explainability research sits inside this community and asks whether a model’s output can be accounted for in terms a person can check. Rudin (2019) argues that high-stakes decisions call for models that are interpretable by construction, since an explanation attached to an opaque model after the fact can misstate what the model actually did, an argument that bears directly on the child protection systems social work scholarship has described as black boxes practitioners cannot interrogate (Gillingham, 2016). AI safety, the newest of these communities and one whose research output circulates largely as preprints from industry laboratories, prioritizes  preventing harm from advanced models through problems of robustness, oversight, and safe deployment (Amodei et al., 2016; Ganguli et al., 2022). The names for this concern have moved in waves, from algorithmic fairness to responsible AI to AI safety, and that instability is one indication that these communities have not yet consolidated as disciplines.

**Table 1**. Disciplines and Research Communities around AI-Era Technology

| Field | Social work connection | Principal community |
|---|---|---|
| Human-computer interaction | Co-design and participatory design of digital services | ACM CHI; ACM CSCW |
| Artificial intelligence and machine learning | Method tutorials translating models for research audiences | AAAI; NeurIPS; ICML; ACL; EMNLP |

| Data science | Predictive analytics and text mining on administrative and case records | ACM SIGKDD (KDD) |
|---|---|---|
| Computational social science | Computational text analysis of social media and large-scale administrative data | International Conference on Computational Social Science (IC2S2); AAAI International Conference on Web and Social Media (ICWSM); Summer Institute in Computational Social Science (SICSS) |
| Information science | The human service informatics lineage | husITa; ASIS&T; the iSchools |
| Science and technology studies | Critical analysis of automated public services, AI's political economy, and its environmental and data-center costs | Society for Social Studies of Science (4S); European Association for the Study of Science and Technology (EASST) |
| Digital health | Co-design, deployment, and evaluation of digital mental health and internet interventions | Digital Medicine Society (DiME); Society for Digital Mental Health (SDMH); International and European Societies for Research on Internet Interventions (ISRII, ESRII) |
| Health informatics | Electronic health records and clinical documentation, including case-note analysis | American Medical Informatics Association (AMIA) |
| Implementation science | Technology adoption in human service organizations | Society for Implementation Research Collaboration (SIRC) |
| Robotics and human-robot interaction | Care robotics with the populations social work serves | ACM/IEEE International Conference on Human-Robot Interaction (HRI) |
| Learning sciences | AI literacy and curriculum | International Society of the Learning Sciences (ISLS) |
| Responsible AI | Fairness auditing of allocation and child welfare models | ACM FAccT; ACM EAAMO; AAAI AIES |
| AI safety | - | Industry laboratories; research circulates as preprints |

*Note.* ACM = Association for Computing Machinery; CHI = Conference on Human Factors in Computing Systems; CSCW = Conference on Computer-Supported Cooperative Work and Social Computing; AAAI = Association for the Advancement of Artificial Intelligence; NeurIPS = Conference on Neural Information Processing Systems; ICML = International Conference on Machine Learning; ACL = Association for Computational Linguistics; EMNLP = Conference on Empirical Methods in Natural Language Processing; SIGKDD = Special Interest Group on Knowledge Discovery and Data Mining; KDD = Conference on Knowledge Discovery and Data Mining; husITa = Human Services Information Technology Applications; ASIS&T = Association for Information Science and Technology; FAccT = Conference on Fairness, Accountability, and Transparency; EAAMO = Conference on Equity and Access in Algorithms, Mechanisms, and Optimization; AIES = Conference on AI, Ethics, and Society.

**Social Workers as Technology Decision-Makers**

The social work profession's response to this extends beyond literacy for practice and ethics for research. Interviews with software engineers found that even relatively powerful engineers often lacked the power to resolve the ethical concerns they held, with personal precarity and organizational incentives limiting what they could act on (Widder et al., 2023). Ethical concern that arrives without decision authority does not settle what a system does. We argue that social workers could and should hold decision-making roles across these settings, with the product manager role as the primary example: its daily work is aligned with the methods social work already trains, and its center is domain judgment about which problem to solve, for whom, and in what order (Cagan, 2019; Maglyas et al., 2013). The product manager role carries no licensure and no single degree pathway. When the 2018 revision of the Standard Occupational Classification was asked to create a detailed product manager occupation, the Standard Occupational Classification Policy Committee declined, on the ground that workers with that title are classified by the work they perform under existing occupations; in consequence, no official employment or wage series exists for the title (National Center for O*NET Development, n.d.; U.S. Bureau of Labor Statistics, Standard Occupational Classification Policy Committee, 2018). The absence of a credential gate is not the same as open access. Where the role sits and how much authority it carries vary with how formally an organization has defined it (Ebert, 2007; Maglyas et al., 2013). Whether MSW holders are hired into these positions, and at what rate, is not documented.

Program manager, public policy or government affairs manager, and social impact manager sit outside the engineering organization. The first has what the product manager title lacks, which is a place in the official occupational classification. Social and community service managers, classified at 11-9151, plan, direct, or coordinate the activities of a social service program or community outreach organization, oversee its budget and its policies on participant involvement, program requirements, and benefits, and may direct social workers, counselors, and probation officers; half of surveyed incumbents report a bachelor's degree and a quarter a master's (National Center for O*NET Development, n.d.-b). The definition turns on the program. Where an organization runs an AI tool through its own staff, as a county does when caseworkers use a benefits assistant and a hotline does when counselors use a crisis tool, someone holds that program. Where a person downloads a self-serve application and operates it alone, no one does, and the role is not there to be held.

European regulation draws the same line and attaches obligations to it. The AI Act distinguishes the provider, who develops a system and places it on the market, from the deployer, who uses it under its own authority, and it excludes use in the course of a personal non-professional activity from the deployer definition (European Union, 2024). The deployer decides which populations a system is turned on for and under what commitments, and the clinical safety officer requirement described above attaches to the deploying organization as well as the manufacturer (NHS England, 2025a, 2025b). A benefits applicant meets an eligibility system when a county turns it on and not when an engineer writes it. Product management corresponds most closely to social work method, and the deployer-side roles reach the release decisions without passing through an engineering hiring process.

Table 2 sets the nine competencies of the 2022 Educational Policy and Accreditation Standards, in their verbatim titles (Council on Social Work Education, 2022), against the product-management work each corresponds to, with the product side drawn from the empirical role literature (Maglyas et al., 2013) and the competency structure practitioners use (Cagan, 2019). Needs assessment, the collaborative definition of a presenting problem with the people who have it, is the method that product discovery research performs under another name, and the field technique discovery most often uses, observation and interviewing in the context of real work, entered product practice from contextual inquiry (Holtzblatt & Beyer, 1997). The remaining correspondences in Table 2 run the same way, from group facilitation to co-design workshops and from program evaluation to product evaluation and metric definition.

**Table 2.** The 2022 EPAS Competencies and Corresponding Product Management Tasks

| EPAS 2022 competency | What the competency trains | Corresponding product-management task |
|---|---|---|
| 1. Demonstrate Ethical and Professional Behavior | Applying an enforceable ethics code and models of ethical decision making to practice situations; managing professional boundaries and judgment | Judging acceptable product behavior before release; writing requirements that operationalize consent, privacy, and safety commitments, translating into product terms the obligations legal counsel sets |
| 2. Advance Human Rights and Social, Racial, Economic, and Environmental Justice | Analyzing how resources, rights, and structural barriers are distributed across populations | Proposing and defending which user problems and which populations the plan serves, within the priorities senior leadership and regulatory obligations set; weighing equity consequences of feature trade-offs |
| 3. Engage Anti-Racism, Diversity, Equity, and Inclusion (ADEI) in Practice | Recognizing how bias, power, and privilege operate in institutions and in the worker's own judgment | Deciding which users research must include; requiring products to be tested for unequal performance across groups |
| 4. Engage in Practice-Informed Research and Research-Informed Practice | Critically appraising evidence; qualitative and quantitative methods; recognizing bias in design and interpretation | Commissioning and appraising discovery research; interrogating experiment results and vendor evidence claims |
| 5. Engage in Policy Practice | Analyzing, formulating, and advocating for policy within practice settings | Anticipating regulatory constraints on the product; working with the public policy, government affairs, and legal |
| | | teams that hold compliance authority |
| 6. Engage with Individuals, Families, Groups, Organizations, and Communities | Relationship building; interviewing, including motivational interviewing; person-in-environment work with clients and constituencies | Conducting user interviews; stakeholder alignment across engineering, design, clinical, and business functions |
| 7. Assess Individuals, Families, Groups, Organizations, and Communities | Needs assessment; collaborative problem definition; ecomap and person-in-environment assessment | Discovery research and problem definition; diagramming a user's route through a service |
| 8. Intervene with Individuals, Families, Groups, Organizations, and Communities | Selecting and implementing evidence-informed interventions; negotiation, mediation, and group facilitation | Deciding what the team builds under constraint; negotiating scope with engineering; running co-design workshops |
| 9. Evaluate Practice with Individuals, Families, Groups, Organizations, and Communities | Program evaluation; outcome measurement with qualitative and quantitative methods | Product evaluation and metric definition; ownership of the build plan and the measures of success; judging whether a released feature achieved its outcome |

*Note.* Competency titles are quoted verbatim from the 2022 Educational Policy and Accreditation Standards (Council on Social Work Education, 2022, pp. 8-12). The product-management column draws on Maglyas et al. (2013) and Cagan (2019). The product-management column names the work the role performs. Formal authority over several of these decisions sits elsewhere in the organization, with legal counsel setting privacy and safety obligations, senior leadership and government affairs setting prioritization, and public policy teams owning regulatory compliance.

Named precedents exist in trust and safety and platform policy. Jerrel Peterson, an MSW-trained professional who began in direct services and Head Start programs, leads global content policy at Spotify after heading safety policy at Twitter/X (Social Work to Wealth, 2025b). Leslie Taylor, an MSW-trained professional who began as an in-home therapist and later reviewed child-exploitation cases at the National Center for Missing & Exploited Children, joined Snap's trust and safety team as its fourth member and now directs trust and safety solutions at Foundever after a senior role at Adobe (Social Work to Wealth, 2025a). Therapists in Tech, a nonprofit of more than 2,000 members supporting mental health professionals moving into technology companies, names social work licensure designations (LICSW, LCSW-BACS, LCSW-S) alongside psychology and counseling credentials in its own description of its membership (Therapists in Tech, n.d.). The Trevor Project's Crisis Contact Simulator, an AI tool for training crisis counselors, was built by the organization's own AI, engineering, and product staff working with Google.org fellows, and the organization's training team defined the human evaluation rubric the tool had to satisfy. The disciplines represented on that team include clinical psychology and education (Ball, 2021; The Trevor Project, 2021).

The five groups in Figure 2 differ in how far they have hardened into named job families. Product and governance titles are advertised, salaried, and increasingly credentialed. Technology leadership in human service organizations often exists as duties without a title, and grantee collaboration is the least named of the five, living inside program management and development work. Product roles decide what gets built and on what evidence, and governance roles decide what may be released and what must change. The AI governance job family is growing, with 77% of surveyed organizations working on AI governance and a certification, the Artificial Intelligence Governance Professional, open to candidates without engineering backgrounds (IAPP & Credo AI, 2025).

Technology leadership roles inside human service organizations decide what the organization adopts and on what terms. For example, in small nonprofits this work has long been carried informally by whoever happened to be nearby, the *accidental techie* (Bennett et al., 2005), and the proposal here is to make that role deliberate. Grantee collaboration roles decide what a collaboration with a technology company owes the organization and its clients. That question is live when a funder embeds its own fellows in grantee organizations, as Google.org (Google.org, n.d.) and Anthropic's Claude Corps (Anthropic, 2026) do, or denominates its awards partly in vendor credits. It is live again when a funder ends donated infrastructure on its own schedule, as Microsoft did when it retired free license tiers relied on by roughly 400,000 nonprofits (Thompson, 2025). Policy and oversight roles decide what the rules are; the social work profession's policy practice competency has a natural application in mechanisms such as Canada's mandatory algorithmic impact assessment for federal automated decision systems (Treasury Board of Canada Secretariat, 2024).

**Figure 2**. Technology decision-related roles of social workers in the age of AI

Policy and oversight roles
decide what the rules and regulations are
AI policy analyst / policy advocates
Public sphere
Technology industry
Product roles
decide what gets built
product manager / UX researcher
Governance roles
decide what may be released
responsible AI lead / trust & safety / AI governance
Human service organization
Technology leadership
what the organization adopts
technology lead / implementation lead
Clients and communities
Grantee collaboration
decide what the collaboration with tech companies owes
program manager / grant writer
Building
make the tool the setting needs
research technology specialist / civic technologist

Note. Solid boxes are the five groups of decision roles developed in this paper. The dashed box marks building tools directly, which is a further possibility rather than an established job family in these settings.

The five groups above concern deciding what other people build, and a further possibility has opened alongside them, which is that social workers build the tools themselves. Producing a working piece of software once required years of training that an MSW program has no room for, and AI coding assistants now supply the part of that training a domain expert was least likely to have. A person can describe in ordinary language what a tool should do and have a model to write and revise the code, a practice commonly called *vibe coding*. Other professions have begun making this argument for their own members. Chow and Ng (2025) argue in a medical education journal that AI-assisted vibe coding turns clinicians from technology adopters into creators of their own teaching tools, and Moore and Tatonetti (2025) describe the same practice as a route into biomedical software development. Qi and colleagues (2026), working at a school of social work, tested whether small language models running on local hardware can match larger commercial models at identifying constructs in child welfare case records, and found the smaller models competitive on the task, a question that matters wherever records cannot be sent to an outside commercial service. Professional schools have begun teaching this work directly. In one term the University of Michigan Law School's AI Law and Policy Clinic placed students with courts, legal aid organizations, and nonprofits, and eight student teams produced working prototypes, among them a tool for assembling court-ready parenting time schedules and a tool for completing delegation of parental authority forms for immigrant families facing detention (University of Michigan Law School, 2026).

The limits of this are as well documented as the possibility. Non-experts asked to direct a language model overestimate how well their instructions convey what they want, and they repair failures by trial and error without a working account of why an attempt failed (Zamfirescu-Pereira et al., 2023). End-user programmers using code-generating models meet a gap between the terms in which they understand their own problem and the terms the model needs, a gap that ordinary programming experience would have closed (Liu et al., 2023). The barrier to producing a working tool has dropped and has not disappeared, and what remains of it is what a curriculum can supply. The case for teaching this material does not rest on the claim that anyone can now build anything. It rests on a narrower one: for a small tool serving a specific agency, the scarce input is knowing which problem is worth solving and what a wrong answer would cost a client, and that is the input an MSW already supplies.

**Production Literacy**

Ahn and colleagues (2025) specified the AI literacy that practice requires, organized around Long and Magerko's (2020) five areas, running from understanding what AI systems are and can do to evaluating how they should be used and how people perceive them. That specification fits its purpose, competent and critical practice around AI systems built by others, and the positions named here require more, which we call *production literacy*. It contains the AI literacy Ahn and colleagues (2025) specify and adds five things to it. A social worker holding one of these positions has to understand how a model was trained and evaluated, decide what a system should do, judge what a vendor is selling, read what a measured result means, and know which rules bind the employer. The dependency runs one way: ethical judgment about a system sharpens only as understanding of the system's mechanism deepens, because a reviewer who cannot say how a model was trained, on what data, and against what evaluation cannot say where its harms will concentrate. Ahn and colleagues (2025) make a version of this point for practice when they argue that understanding how AI works is what lets a practitioner question the representativeness of a model's training data. The argument extends upward, because the roles named here ask the same questions with decision authority attached.

The additions would benefit from being produced in a specific order. A technical understanding of past tool use could come first, meaning how models are trained and evaluated, where their failure modes lie, and what a claimed accuracy figure does and does not establish. The methods tutorials already noted supply entry points at this level (B. E. Perron, Rivenburgh, et al., 2025; B. Perron et al., 2026). How products are discovered, versioned, and changed could build on technical understanding (Boehm, 1981; Cagan, 2019), without which a social worker in a product organization cannot convert judgment into requirements. How vendors price and withdraw what they sell could come next, since human service organizations mostly buy technology; understanding why a vendor's plan does not bend to one agency, what a donated license is worth when it can be withdrawn (Thompson, 2025), and which contract terms preserve the ability to switch vendors (J. Y. Kim & Kesari, 2026) is a subject of its own. How product teams judge whether something worked would follow, since they reason through controlled experiments with live users and movements in defined metrics (M. Kim et al., 2016). That reasoning is nearer to program evaluation than it seems, since an A/B test assigns users at random to concurrent versions and compares a defined outcome, and the differences are that the outcomes are chosen for the business and the follow-up window is short. A social worker trained in program evaluation must learn where the translation holds and where a metric that moved says nothing about whether a client was helped. Which rules bind an employer and which only advise would be last: the European Union's AI Act is binding regulation with a risk-tiered structure (European Union, 2024), while the NIST AI Risk Management Framework is voluntary guidance (Tabassi, 2023), and confusing the two misreads what an employer's compliance obligations actually are. The MIT AI Risk Initiative maintains a public reference for these terms, synthesizing published AI risk taxonomies into one classification alongside a catalogue of AI governance documents, which gives a social worker entering a governance role a list of the harms the field already names (Slattery et al., 2026).

None of this would require a second master's degree. Instead, the additional training could be acquired through structured coursework inside existing MSW programs, through civic technology volunteering that supplies product experience on public-interest problems in the organizations whose postings already prefer service-domain backgrounds, and through certifications such as the AIGP that do not presuppose an engineering background (IAPP, n.d.).

Curricular space is a constraint, since an accredited program already carries the nine competencies within a fixed number of credit hours (Council on Social Work Education, 2022), and continuing education is the route that does not compete for that space. Nearly every jurisdiction that licenses social workers requires documented continuing competence for renewal (Association of Social Work Boards, 2021), so a continuing education sequence in production literacy would reach practitioners already in the field without waiting for a curriculum revision. The routes also compound, since a student who has carried one civic technology discovery project from interviews through problem definition to metric selection has practiced several of the additions at once, which is one reason placement-based routes may outperform coursework alone. Students themselves appear ready for this framing, as interviews with MSW students found them asking for concrete preparation to work with and on these systems (Kapur et al., 2026).

### Ethics for the Deployment Era

The social work profession's ethics for technology govern the practitioner's own use of tools, and they do so in detail. The NASW Code of Ethics addresses informed consent for technology-mediated services, including verifying identity and capacity and offering alternatives (Standard 1.03[e]-[i]); competence in using technology and complying with the laws of the jurisdictions involved (1.04[d]-[e]); differences among clients in access to and comfort with technology (1.05[e]); confidentiality safeguards for electronic communication, breach notification, and electronic searches of clients (1.07[m]-[r]); and technology-mediated access to records (1.08[b]) (National Association of Social Workers, 2021). Standard 1.07(m) is a case in point: it directs the practitioner to use safeguards such as encryption and passwords, technical language written for the person using a system, with no counterpart standard for the people who decide what a system encrypts, retains, or shares. The joint standards issued by NASW, ASWB, CSWE, and the Clinical Social Work Association organize 55 technology standards across the provision of public information, service design and delivery, information management, and education and supervision (NASW et al., 2017). Every standard addresses a social worker using technology, and none addresses a social worker building, procuring, or governing it. Lee and colleagues (2026) have supplied an ethics for researchers' own use of AI; what remains understudied is the deployment stage, the ethics of the decisions made about systems before any practitioner uses them.

The content of the deployment ethics is already visible in the decisions themselves. Procurement is one such decision: contract terms settle auditability, data rights, and exit, and guidance now exists that a practice-trained reader can evaluate (J. Y. Kim & Kesari, 2026). Data governance is another, since automated eligibility and risk-scoring systems have concentrated surveillance and error on poor families in documented cases (Eubanks, 2018). Accountability mechanisms carry their own ethics: an algorithmic impact assessment can define impacts in ways convenient to the organization and distant from the harms people experience (Metcalf et al., 2021), yet where governments mandate such assessments, as Canada does for federal automated decision systems (Treasury Board of Canada Secretariat, 2024), someone must write them, and the social work profession's needs-assessment and harm-evaluation training transfers directly to that writing. The frontline is an ethical site too, since when deployed systems fail, responsibility tends to migrate to the practitioner nearest the failure. Garrett (2025) argues, borrowing Elish's term, that AI documentation tools in statutory practice position the social worker as a moral crumple zone, absorbing accountability for outputs no one at the front line controls. The practitioner is the person who notices harm first; deployment ethics asks who was positioned to prevent it.

An integrative review of the field's AI literature from 2020 to 2024 found it heavy on anticipated benefits, thin on ethics beyond downstream checklists, and largely silent on AI's political economy, labor conditions, and environmental costs, concluding that the field's writing treats AI as an area of computation and overlooks the political, financial, and social system around it (Massey et al., 2026). If that diagnosis is right, more commentary written at a distance from those settings will not supply the remedy; structurally trained people inside them might. The critique also carries an honest warning for our own proposal, since employment in the industry could deepen the optimism Massey and colleagues diagnose; whether it does is an empirical question.

Further, Young and colleagues (2022) documented corporate entanglement at the field's flagship accountability conference and argued that remedies should be drawn from conflict-of-interest management in licensed professions such as medicine and public health. Their argument cautions that governance work inside companies can become legitimation, the pattern the ethics-washing literature describes, in which voluntary ethics language substitutes for binding constraint (Wagner, 2019), though Bietti (2020) warns with equal force against dismissing all industry ethics work as performance. It also names the asset a licensed profession brings: obligations that attach to an external body and persist across employers. The remedies Young and colleagues (2022) propose, disclosure requirements and independent participation by affected communities among them, presuppose actors whose obligations run beyond their employer, and the licensed professions they point to as models are professions of exactly social work's kind. A social worker's accountability to a licensing board and an enforceable code does not dissolve at a job change, which is the external attachment the capture critique finds missing, and the 2022-2023 layoffs that thinned AI ethics and trust and safety teams (Field & Vanian, 2023) show why an ethics worker whose standing lives entirely inside one employer is structurally exposed.

**Implications and a Research Agenda**

For education, the accounts given in this paper are teachable now. The account of the technology team and its decisions, together with the disciplines shown in Figure 1 and Table 1 and the five role groups in Figure 2, supplies a syllabus outline for existing technology and macro practice courses, and the production literacy additions name the units such courses currently lack, alongside the AI literacy content the field has already specified (Ahn et al., 2025; Rodriguez et al., 2024). An AI competency for a future EPAS revision has been proposed elsewhere (Rodriguez et al., 2024). The correspondences in Table 2 suggest that the existing nine competencies already carry the transferable methods. Field education is another route, as placements in civic technology organizations, AI governance functions, and human service technology leadership would give students supervised hours inside these settings, and a clinic on the law school model, in which supervised student teams build tools for community partners inside a single term, would put the same work in the curriculum itself (University of Michigan Law School, 2026), extending the technology field placements proposed in earlier work (Mathiyazhagan, 2022). Building such placements requires field instructors who know these settings, and the small population of MSW holders already working in them, visible through communities such as Therapists in Tech, is the natural first recruitment pool.

For the social work profession's institutions, the comparison case is nursing. Nursing informatics was recognized as a nursing specialty in 1992 and first certified in 1995, and the credential, renamed Nursing Informatics-Board Certified in 2023, is active today through the American Nurses Credentialing Center, the product of institutional investment sustained across three decades (*American Nurses Credentialing Center*, n.d.; Nashwan et al., 2025). Social work has no equivalent while practitioner surveys report social workers asking for training and oversight infrastructure of exactly this kind (Pandya, 2026). NASW's own published credential list spans addictions, case management, clinical practice, education, gerontology, health care, hospice, military service, and youth and family practice, and contains no credential naming technology, informatics, or AI. Psychology has not built a dedicated credential either, and its most recent technology-facing guideline revision is scoped to telehealth delivery (American Psychological Association, 2024).

The typology we propose should be treated as a set of testable claims. Job posting analyses could track which technology decision roles name social work credentials and how that changes over time, and career-history studies could trace how MSW holders enter and advance in them. Case studies could measure how much decision authority embedded social workers actually hold, and whether industry employment weakens the critical stance the social work profession brings (Massey et al., 2026). Further work could evaluate whether industry-structure teaching and technology field placements change students' competencies and destinations and compare how nursing informatics built its credential while psychology gated clinical product and AI safety roles.

## Conclusion

The technology sector is building and operating the systems that run inside the domains social work serves. Social workers carry the domain judgment those systems act on, and they are rarely present when the decisions that set a system's behavior are made. This paper argued that they should hold those decision roles themselves, across the technology industry, human service organizations, and policy institutions, with product management as the primary example because its methods correspond to social work training. Whether that correspondence holds in hiring, in authority, and in career trajectories is answerable by empirical study, and the credentialing comparison with nursing suggests the institutional work should not wait for the answer.